\documentclass{acmart}
\usepackage{CJKutf8}
\usepackage{tabularx}
\usepackage{threeparttable}

\AtBeginDocument{%
  }

\setcopyright{acmlicensed}
\copyrightyear{2018}
\acmYear{2018}
\acmDOI{XXXXXXX.XXXXXXX}

\acmConference[Conference acronym 'XX]{Make sure to enter the correct
  conference title from your rights confirmation emai}{June 03--05,
  2018}{Woodstock, NY}
\acmISBN{978-1-4503-XXXX-X/18/06}

\begin{document}


\title{An Empirical Study of ChatGPT-Related Projects and Their Issues on GitHub}

\author{Zheng Lin}
\orcid{0009-0006-6508-2651}
\affiliation{%
  \institution{School of Software Engineering, Sun Yat-sen University}
  \city{Zhuhai}
  \country{China}}
\email{linzh228@mail2.sysu.edu.cn}

\author{Neng Zhang}
\orcid{0000-0001-8662-5690}
\affiliation{%
  \institution{School of Software Engineering, Sun Yat-sen University}
  \city{Zhuhai}
  \country{China}}
\email{zhangn279@mail.sysu.edu.cn}
\authornote{Corresponding author.}

\author{Chao Liu}
\affiliation{%
  \institution{School of Big Data \& Software Engineering, Chongqing University}
  \city{Chongqing}
  \country{China}}
\email{liu.chao@cqu.edu.cn}

\author{Zibin Zheng}
\affiliation{%
  \institution{School of Software Engineering, Sun Yat-sen University}
  \city{Zhuhai}
  \country{China}}
\email{zhzibin@mail.sysu.edu.cn}

\renewcommand{\shortauthors}{Zheng et al.}

\begin{abstract}
  Due to its powerful capabilities in natural language understanding and content generation, ChatGPT has received widespread attention since its launch in 2022. An increasing number of ChatGPT-related projects (that enhance the capabilities of ChatGPT, develop applications by calling ChatGPT APIs, improve the usage of ChatGPT, etc.) are being released on GitHub and have sparked widespread discussions. However, GitHub does not provide a detailed classification of these projects to help users effectively explore interested projects. Additionally, the issues raised by users for these projects cover various aspects, e.g., installation, usage, and updates. It would be valuable to help developers prioritize more urgent issues and improve development efficiency. Unfortunately, there is currently no research focused on understanding the categories and issues of ChatGPT-related projects. To fill this gap, we retrieved 71,244 projects from GitHub using the keyword `ChatGPT' and selected the top 200 representative projects with the highest numbers of stars as our dataset. By analyzing the project descriptions, we identified three primary categories of ChatGPT-related projects, namely \textit{ChatGPT Implementation \& Training}, \textit{ChatGPT Application}, \textit{ChatGPT Improvement \& Extension}. Next, we applied a topic modeling technique to 23,609 issues of those projects and identified ten issue topics, e.g., \textit{model reply} and \textit{interaction interface}. We further analyzed the popularity, difficulty, and evolution of each issue topic within the three project categories. Our main findings are: 1) The increase in the number of projects within the three categories is closely related to the development of ChatGPT; and 2) There are significant differences in the popularity, difficulty, and evolutionary trends of the issue topics across the three project categories. Based on these findings, we finally provided implications for project developers and platform managers on how to better develop and manage ChatGPT-related projects, such as offering more fine-grained tags to categorize projects to facilitate their exploration.
\end{abstract}



\received{20 February 2007}
\received[revised]{12 March 2009}
\received[accepted]{5 June 2009}

\maketitle

\section{Introduction}

In recent years, with the rapid development of artificial intelligence (AI), a considerable number of chat robots~\cite{Olujimi2023,Zhang2022,Song2023,Sara2021} (also referred to as chatbots or conversational agents) have been developed. After receiving natural language input, chatbots can generate responses in various forms such as text, code, images, and video~\cite{Meredith2023,Crawford2021,Pablo2022,Dagkoulis2023}. Among these chatbots, ChatGPT\footnote{\url{https://chat.openai.com/}} released by OpenAI on November 30, 2022 has received widespread attention. It is supported by the GPT-3.5 large language model (LLM). After its emergence, ChatGPT quickly gained popularity, surpassing one million users within five days and exceeding 100 million users within two months~\cite{Firat2023WhatifGPT4BecameAutonomous,Firat2023WhatChatGPTmeansforuniversities}. ChatGPT has numerous outstanding capabilities. For example, ChatGPT can capture contextual information from conversations and generate responses in the form of human conversation, greatly enhancing the interactive experience for users~\cite{Lund2023}. ChatGPT is also creative. For instance, when a programmer expresses his/her functional requirements to ChatGPT and specifies the target programming language, ChatGPT can automatically generate the corresponding source code~\cite{Liu2023}. In March 2023, OpenAI released an upgraded version GPT-4, which further enhanced the capabilities of ChatGPT~\cite{OpenAI2023}.

Since the release of ChatGPT, a large number of ChatGPT-related projects (that enhance the capabilities of ChatGPT, develop applications by calling ChatGPT APIs, improve the usage of ChatGPT, etc.) have been increasingly created and hosted on GitHub, the largest open-source project hosting platform worldwide. These projects have also attracted a great number of discussions. By investigating the types of these emerging projects, it is possible to provide more fine-grained tags for the GitHub platform to categorize projects and enhance the project management capability. In addition, studying the popularity, difficulty, and evolutionary trends of issues raised by users can help developers prioritize key issues and improve development efficiency. However, there is no comprehensive study on investigating the types of these projects and their associated issues. 
To fill this gap, we collected 71,244 projects 
from GitHub using the keyword `ChatGPT’ and selected the top 200 projects with the highest number of stars. 
Using these representative projects, we conducted several analyses to answer the following research questions (RQs). 

\textbf{RQ1. What categories of ChatGPT-related projects have been developed?}

We manually examined the descriptions of 200 projects and identified three primary categories of ChatGPT-related projects, namely \textit{ChatGPT Implementation \& Training}, \textit{ChatGPT Application}, \textit{ChatGPT Improvement \& Extension}, which include 8, 55, and 55 projects, respectively. Among the remaining 82 projects, 63 projects are irrelevant to ChatGPT (as explained in Section~\ref{subsec:project categorization}) 
and 19 projects belong to other minor categories, e.g., a collection of ChatGPT mirror websites or tools.
The categories can help to understand how developers integrate ChatGPT into their projects and are also valuable for GitHub in improving the management and exploration of ChatGPT-related projects, as detailed in Section~\ref{subsec:implications}. 

\textbf{RQ2. What are the topics of issues discussed in ChatGPT-related projects?}

We applied the Latent Dirichlet Allocation (LDA)~\cite{Blei2003} topic modeling technique to a set of 23,609 issues collected from 118 ChatGPT-related projects belonging to three primary categories. Since there is a significant portion (48.5\%) of issues containing Chinese words, we propose a method to train a unified LDA model for issues expressed in both English and Chinese by semantically aligning the two kinds of words (see Sections~\ref{subsec:issue collection and preprocess} and~\ref{subsec:train LDA}), which results in ten issue topics,  
such as \textit{interaction interface} and \textit{gpt conversation}.
We further analyzed the popularity of these issue topics within each project category. The results show that the popularity of each issue topic exhibits a significant difference across different categories. These issue topics and their popularity provide a comprehensive understanding of the problems that users encountered when using ChatGPT-related projects and also enable developers to improve the quality of their projects by prioritizing popular issues. 

\textbf{RQ3. How difficult is it to address the issues of each topic within different project categories?}

We analyzed the difficulty of addressing the issues of each topic within the three project categories based on the average attention (i.e., the average number of comments and participants) received by the issues and closing rate of the issues. The results show significant differences across different project categories. For example, in the \textit{ChatGPT Implementation \& Training} category, although the \textit{model reply} topic receives high attention, its closing rate is the lowest (57.2\%), indicating that the issues of this topic are difficult to solve. In contrast, in the \textit{ChatGPT Improvement \& Extension} category, the closing rate of \textit{model reply} is the highest (92.2\%). Understanding the difficulty of issue topics, along with their popularity, can help developers better prioritize popular and challenging issues in their projects.

\textbf{RQ4. How do the issue topics evolve over time within different project categories?}

We measured evolution trends of the issue topics using the Mann-Kendall trend test~\cite{Shourov2019}. The issue topics show different evolution trends across the three project categories. For instance, the \textit{npm runtime error} topic shows an increasing trend in the \textit{ChatGPT Implementation \& Training} category and a decreasing trend in the \textit{ChatGPT Improvement \& Extension} category, while there is no significant trend in the \textit{ChatGPT Application} category. Based on the evolution trends of issue topics, developers can respond more quickly to changes in user requirements and adjust future development directions in a timely manner.

In summary, the main contributions of this study are listed as follows. 

\begin{itemize}
\item We manually identified three primary categories of open-source projects closely related to ChatGPT on GitHub. The categories can be used to better organize ChatGPT-related projects on GitHub or other project hosting platforms (e.g., Gitee\footnote{\url{https://gitee.com/}}) and help users search for desired projects more effectively.
\item We proposed a method for aligning Chinese words with English words and thus simultaneously building topic models for the issues of ChatGPT-related projects (or other text corpus) that are written in both English and Chinese.
\item We discovered ten topics from the issues of ChatGPT-related projects and further analyzed the popularity, difficulty, and evolution of these issue topics within different project categories. The results can facilitate the understanding and handling arrangements of the issues raised for ChatGPT-related projects.
\item We provided suggestions for developers and open-source platforms to improve the development and management of ChatGPT-related projects.
\end{itemize}

The rest of this paper is structured as follows. Section~\ref{sec:methodology} describes our research methodology. Section~\ref{sec:results} presents answers to the four RQs. Section~\ref{sec:discussion} provides implications for project developers and managers based on our findings and discusses threats to the validity of our research. Section~\ref{sec:related work} reviews related work. Section~\ref{sec:conclusion} concludes the paper and discusses future work.

\section{Research Methodology} \label{sec:methodology}

\begin{figure}
  \centering
  \includegraphics[width=\textwidth, trim = 0 0 0 0, clip]{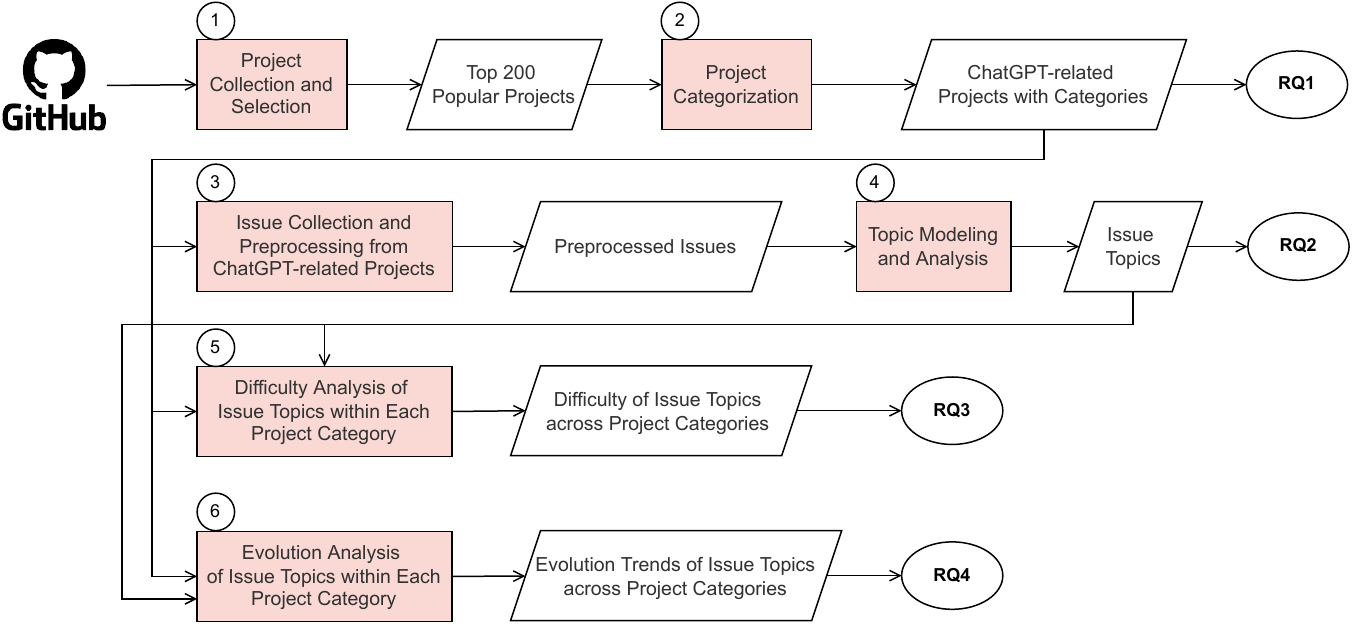}
  \caption{Our research methodology}
  \label{fig:research methodology}
\end{figure}

Figure~\ref{fig:research methodology} shows an overview of our research methodology. At first, we retrieved an initial set of open-source projects that might be relevant to ChatGPT from GitHub. For each project, we collected various attributes, e.g., the project description, number of stars, and number of issues. We filtered projects without issues and selected the top 200 projects with the highest numbers of stars as our research dataset. By manually examining the project descriptions, we identified and classified ChatGPT-related projects into three primary categories, answering RQ1. We further collected and preprocessed the issues from these ChatGPT-related projects. Next, we trained a topic model to discover the topics of issues, answering RQ2. After that, by measuring the attention and the closed status of issues, we analyzed the difficulty in addressing issues across different topics, answering RQ3. Finally, we analyzed the evolution trend of each issue topic over time within the three categories, answering RQ4.

\subsection{Project Collection and Selection} \label{subsec:project collection}

To conduct our study, we needed to collect a set of representative projects related to ChatGPT. As the world's largest open-source project platform, GitHub has attracted a large number of ChatGPT-related projects, and it offers a series of web APIs\footnote{\url{https://api.github.com/}} to facilitate the retrieval of various kinds of data, e.g., projects and their issues.

We initially retrieved the projects created from Nov. 30, 2022 (the release date of ChatGPT) to Oct. 14, 2023, using the GitHub API with the search keyword `ChatGPT', which resulted in 71,244 projects. For each project, we collected several attributes, including the full name, number of issues, number of stars, created time, and description. Specifically, the description of a project was extracted from the `About' section and the README file, which summarize some important details, such as research intent, functionalities, implementation methods, and usage. 
Table~\ref{tab:project attribute} presents a brief explanation of these attributes. 

We observed that a large number of the collected projects do not have any issues. Since one of our key objectives is to identify potential user requirements through the issues of ChatGPT-related projects, we removed the projects without any issues as they cannot provide valuable information. 
Table~\ref{tab:statistical indicators of issues number} presents several statistics on the number of issues of the retained 4,541 projects. As can be seen, these projects have less than 16 issues on average; 50\% projects have only one or two issues; and 75\% projects have no more than eight issues. 
The projects with only a small number of issues may not be of much interest to users and may be unsuitable for our study. Generally, the number of stars could reflect the popularity of a project~\cite{Cohen2018}. Popular projects often have more issues reported by users. To gather a set of representative projects with high popularity, we sorted the 4,541 projects by the number of stars in descending order and selected the top 200 projects as our research dataset. In total, the top 200 projects have 39,985 issues, surpassing half of the total number of issues of the 4,541 projects. Table~\ref{tab:statistical indicators of issues number} also presents several statistics calculated for the number of issues of these 200 projects.

\subsection{Project Categorization} \label{subsec:project categorization}

Based on our analysis of several projects, there are different types of projects about ChatGPT. For example, some projects propose methods to improve the capability or efficiency of ChatGPT, and some others focus on applying ChatGPT in specific task scenarios, e.g., code generation. 

To help researchers, developers, and users understand what primary types of projects have been developed by leveraging ChatGPT, we decided to categorize the top 200 popular projects collected in the previous step. 
Since there were no pre-defined categories, the first and second co-authors of this paper performed the categorization task using an open card sorting approach in two phases~\cite{Gao2023}. In the first phase, both co-authors manually and independently examined the description of each project and built a set of categories. After that, they discussed the categories together, which resulted in a common set of five categories: \textit{ChatGPT Implementation \& Training}, \textit{ChatGPT Application}, \textit{ChatGPT Improvement \& Extension}, \textit{Other ChatGPT-related Project}, and \textit{Irrelevant Project}. The former three categories include the projects developed using ChatGPT. Their explanations can refer to Section~\ref{subsec:RQ1}. The \textit{Other ChatGPT-related Project} category includes the projects relevant to ChatGPT but do not directly use ChatGPT, e.g., a project offering a collection of ChatGPT website mirrors or tools. The \textit{Irrelevant Project} category includes the projects irrelevant to ChatGPT. Through our analysis, we found that although all the projects were retrieved using the keyword `ChatGPT', a number of projects are not really relevant, but are instead similar to ChatGPT. Such projects were retrieved due to the `chatgpt' tag inaccurately assigned to them. For example, the project `OpenLMLab/MOSS'\footnote{\url{https://github.com/OpenLMLab/MOSS}} develops an open-source tool-enhanced conversational language model similar to ChatGPT, but the developer assigned the `chatgpt' tag to it.

\begin{table}
  \caption{Attributes collected for GitHub projects}
  \label{tab:project attribute}
  \begin{tabularx}{\textwidth}{lX@{}}
    \toprule
    Attribute & Explanation\\
    \midrule
    Full name & A unique identifier of a project, including the owner name and repository name, e.g., `chatanywhere/GPT\_API\_free' \\
    Number of issues & The number of open/closed issues owned by a project\\
    Number of stars & The number of stars owned by a project, which can reflect the popularity of the project\\
    Created time & The created time of a project\\
    Description & The content extracted from the `About’ section and the README file of a project, which summarizes the research intent, functionalities, implementation, and/or usage of the project\\
  \bottomrule
\end{tabularx}
\end{table}

\begin{table}
  \caption{Statistics calculated for the numbers of issues of two sets of GitHub projects}
  \label{tab:statistical indicators of issues number}
  \begin{tabular}{lccccccc}
    \toprule
    Projects & \#Total issues & Min & Max & Mean & Q1 & Q2 (Media) & Q3 \\
    \midrule
    4,541 projects with issues & 72,510 & 1 & 1,776 & 15.97 & 1 & 2 & 8\\
    200 projects with the highest numbers of stars & 39,985 & 1 & 1,776 & 199.93 & 47 & 106.5 & 226.5\\
  \bottomrule
  \end{tabular}
\end{table}

In the second phase, based on the determined categories, the two co-authors independently classified each project into these categories. There were 26 projects for which the co-authors assigned different categories. We measured the inter-rater agreement between the results of both co-authors using the Fleiss Kappa~\cite{Fleiss1971}. The Kappa value was 0.82, which indicated almost perfect agreement. After discussing the disagreements together, both co-authors reached a consensus. There are 8, 55, 55, 19, and 63 projects assigned to the five categories, respectively.

\subsection{Issue Collection and Preprocessing from ChatGPT-related Projects} \label{subsec:issue collection and preprocess}

\subsubsection{Issue Collection}
To investigate what kinds of issues have been reported in the projects developed using ChatGPT, we further collected the issues of the 118(=8+55+55) ChatGPT-related projects belonging to the three primary categories, i.e., \textit{ChatGPT Implementation \& Training}, \textit{ChatGPT Application}, and \textit{ChatGPT Improvement \& Extension}. In total, we obtained 23,609 issues. For each issue, we collected its numerical identifier, title, labels, body, created time, number of comments, number of participants, and state, as listed in Table~\ref{tab:issue attribute}.

\begin{table}
  \caption{Attributes collected for the issues of ChatGPT-related projects}
  \label{tab:issue attribute}
  \begin{tabularx}{\textwidth}{lX@{}}
    \toprule
    Attribute & Explanation\\
    \midrule
    Numerical identifier & A unique identifier of an issue\\
    Title & The title of an issue, which briefly summarizes the issue\\
    Labels & A few keywords or phases (e.g., Bug and Enhancement) of an issue, which are defined by GitHub for issue classification and management\\
    Body & A detailed description of an issue, offering specific content and contextual information about the issue, e.g., the related code, operations, and results\\
    Created time & The created time of an issue\\
    Number of comments & The number of comments associated with an issue\\
    Number of participants & The number of participants who have commented on an issue\\
    State & The state, i.e., open or closed, of an issue\\
  \bottomrule
\end{tabularx}
\end{table}

\subsubsection{Issue Preprocessing}
In this study, we decided to utilize the popular LDA topic modeling technique to discover the discussion topics of issues about ChatGPT-related projects. The main content of an issue is described in its title and body. However, through observation, there are several problems with the content, which may affect the quality of topic modeling. Specifically, (1) there are various kinds of noise, such as code snippets, bug reports, logs, HTML tags, and URLs. (2) Issues are written primarily in either Chinese or English. (3) There are stopwords (e.g., `a' and `the`) and morphologically varied words, e.g., `train' and `training'. Therefore, before applying the LDA technique to the issues, we preprocessed the title and body of each issue using the following five steps. 
Tables \ref{tab:English issue text} and \ref{tab:Chinese issue text} present the preprocessing results of two issues, an English issue\footnote{\url{https://github.com/C-Nedelcu/talk-to-chatgpt/issues/116}} and a Chinese issue\footnote{\url{https://github.com/fuergaosi233/wechat-chatgpt/issues/661}}.

{\bfseries Step 1: Noise content removal.} 
When reporting an issue, the user often attach code snippets, bug reports, or logs to clarify the related program and its execution results or errors. Although these information can help developers understand the issues, our experiments showed that retaining such details often results in generating topics with many meaningless keywords (e.g., `line’, `java’, `py’, `python’, and `package’). 
Thus, similar to previous work~\cite{Izadi2022,Manuel2021,Han2020}, we removed these information using regular expressions based on the markdown syntax of issues. Moreover, we removed several other kinds of noise, including HTML tags (e.g., `<p>'), URLs, file paths, default labels of the markdown-style issue templates (e.g., [BUG] and [FEAT]), numbers, punctuation marks, and non-Chinese/English characters.

{\bfseries Step 2: Word segmentation and lemmatization.} Our analysis showed that a large proportion (48.5\%) of issues containing Chinese words. We used the \verb|posseg| module of jieba\footnote{\url{https://github.com/fxsjy/jieba}} to segment such issues. For the issues that only contain English words, we segmented them using the \verb|word_tokenize| module of NLTK\footnote{\url{https://www.nltk.org}}. In addition, to reduce the interference of grammatical forms of English words when translating Chinese words to English words in step 4, we employed the \verb|WordNetLemmatizer| module of NLTK to convert English words into their basic forms (aka. lemma)~\cite{Miller1995,Manuel2021}. For instance, the lemmas of `creating' and `created' are both `create'.

\begin{table}
  \caption{Preprocessing of an English issue}
  \label{tab:English issue text}
  \begin{tabularx}{\textwidth}{@{}lX@{}}
    \toprule
    Preprocessing process & Result content\\
    \midrule
    Raw text & 
    Widget obstructs the edit button 
    I love this extension, but the widget hovers exactly over the prompt edit button:
    <img width="90" alt="image" src="https://github.com/C-Nedelcu/talk-to-chatgpt/assets/64134453/57b7306c-450c-4d01-8a5f-64033389db42">
    This wouldn't be a big issue normally, but since dragging the widget around behaves very strangely, this is quite annoying.
    Regards!\\
    After step 1 & 
    widget obstructs the edit button i love this extension but the widget hovers exactly over the prompt edit button this wouldn t be a big issue normally but since dragging the widget around behaves very strangely this is quite annoying regards\\
    After steps 2-3 & widget; obstructs; edit; button; love; extension; widget; hovers; prompt; edit; button; wouldn; big; issue; drag; widget; behaves; strangely; annoy; regard;\\
    After steps 4-5 & widget; obstruct; edit; button; love; extens; widget; hover; prompt; edit; button; wouldn; big; issu; drag; widget; behav; strang; annoy; regard;\\
  \bottomrule
\end{tabularx}
\end{table}

\begin{table}
  \caption{Preprocessing of a Chinese issue}
  \label{tab:Chinese issue text}
  \begin{tabularx}{\textwidth}{@{}lX@{}}
    \toprule
    Preprocessing process & Result content\\
    \midrule
    Raw text & 
    \begin{CJK}{UTF8}{gbsn}
    登录问题 ![image](https://user-images.githubusercontent.com/103298985/221082393-591af1e2-82de-4c09-b9f2-e984568a0c2b.png)
    给了正确的账号和密码登录不了，是不是要挂打码平台？，能不能给一个手动登录的？
    \end{CJK}\\
    After step 1 & 
    \begin{CJK}{UTF8}{gbsn}登录问题\end{CJK}
    \ 
    \begin{CJK}{UTF8}{gbsn}给了正确的账号和密码登录不了\end{CJK}
    \ 
    \begin{CJK}{UTF8}{gbsn}是不是要挂打码平台\end{CJK}
    \ 
    \begin{CJK}{UTF8}{gbsn}能不能给一个手动登录的\end{CJK}\\
    After steps 2-3 & 
    \begin{CJK}{UTF8}{gbsn}登录;正确;账号;密码;登录;挂;码;平台;手动;登录;\end{CJK}\\
    After steps 4-5 & login; correct; account; password; login; hang; code; platform; manual; login;\\
  \bottomrule
\end{tabularx}
\end{table}

{\bfseries Step 3: Stopword  removal.} There are stopwords in both Chinese and English that can prevent LDA from identifying meaningful topics~\cite{Manning2008}. We removed stopwords using the Chinese stopword list\footnote{\url{https://github.com/goto456/stopwords}} and the Mallet's English stopword list\footnote{\url{https://github.com/mimno/Mallet/blob/master/stoplists/en.txt}}. These two stopword lists are widely used in previous work~\cite{Ai2024,Cheng2023,Gao2023,Johannes2018}.

{\bfseries Step 4: Chinese word translation.} To uniformly build a topic model for issues expressed in both English and Chinese, we proposed a method for translating Chinese words to semantically equivalent English words. We first collected all Chinese words and all English words from the issues and sorted these two sets of words in descending order by their frequencies, which resulted in a Chinese word list, $CW$, and an English word list, $EW$. Since the Baidu Translation APIs\footnote{\url{https://fanyi-api.baidu.com/product/11}} provide mature documentation for developers and have good performance in translating contextual words~\cite{Xie2022}, we used the Baidu Translation APIs to translate each Chinese word $w \in CW$ to its corresponding English word, $w'$. 
If $w'$ exists in $EW$, we replaced $w$ with $w'$ in all issues.

{\bfseries Step 5: Stemming.} Stemming can effectively handle morphological variations of words and improve the quality of issue texts~\cite{Zhang2019}. For instance, `create', `creation', `creating', and `created' are all stemmed to `creat'. Therefore, similar to previous work~\cite{Gao2023,Silva2021,Zhang2019}, we use the \verb|SnowballStemmer| module of NLTK to reduce each English word to its root form (aka. stem), thereby eliminating morphologically varied words.

\subsection{Topic Modeling and Analysis} \label{subsec:train LDA}

We wanted to discover the discussion topics of issues in ChatGPT-related projects. Although GitHub allows users to annotate issues with some pre-defined labels or tags (e.g., Bug and Enhancement), these labels are often coarse-grained and cannot support fine-grained analysis~\cite{Han2020}. For example, issues labeled as `bug' may be related to API calls, login, pre-training, etc. Moreover, users may not always label issues consistently or correctly~\cite{Siddiq2023}. Therefore, we employed the well-known Latent Dirichlet Allocation (LDA)~\cite{Blei2003} topic modeling technique to discover fine-grained topics from the issues of ChatGPT-related projects. For a corpus of textual documents, LDA assumes that each document has a probability distribution over a set of latent topics, and each topic has a probability distribution over the set of words in the corpus. It uses a bag-of-words model to estimate the topic distribution of each document and the word distribution of each topic based on the co-occurrence of words in the documents~\cite{Barua2014}.

In LDA, there is a key parameter, namely the number of latent topics (denoted as $K$), which controls the granularity of the discovered topics~\cite{Barua2014}. A larger or smaller $K$ may both result in poor performance. To determine the optimal $K$, we utilized the coherence score to evaluate the quality of the trained topic model. The coherence score measures the semantic consistency between words within a topic~\cite{Newman2010}. A higher coherence score indicates stronger semantic associations among words, that is, the topic is more understandable and meaningful.

\begin{figure}
  \centering
  \includegraphics[width=.75\linewidth]{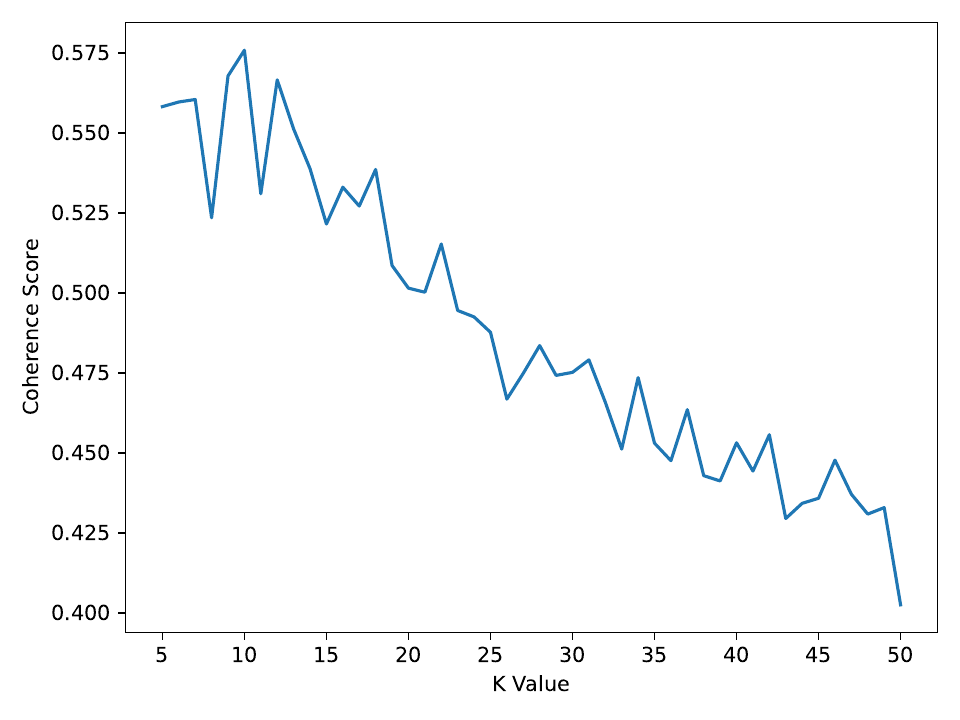}
  \caption{The coherence scores corresponding to different numbers of topics, i.e., $K$.}
  \label{fig:execution_coherence_score}
\end{figure}

We applied the \verb|LdaModel| module provided in the Gensim library\footnote{\url{https://radimrehurek.com/gensim/}} to the entire set of preprocessed issues. We trained different topic models by setting $K$ from 5 to 50 with a step 1. Based on the coherence scores of the models measured using the \verb|CoherenceModel| module (see Fig.~\ref{fig:execution_coherence_score}), we identified the optimal $K=10$ and selected the corresponding model as the final topic model. The first and second co-authors collaboratively named the topics based on the top ten keywords of each topic and some issues belonging to the topic. The topics (e.g., \textit{request method} and \textit{model reply}) with their top ten keywords are presented in Table~\ref{tab:topics of issue texts}. 

According to the trained topic model, each issue possesses a probability distribution of $K$ topics. We denoted the probability of a specific topic $z_k (k=1,...,K)$ in the issue $d_i$ as $\theta(d_i, z_k)$, where $0\leq\theta(d_i, z_k)\leq 1$ and $\sum_{i=1}^{k}\theta(d_i, z_k)=1$. To answer RQ2, we defined the topic with the highest probability as the \textit{dominant topic} of $d_i$, i.e., $dominant\_topic(d_i) = z_k:\theta(d_i,z_k)=\max(\theta(d_i, z_j)), 1\leq j\leq K$. Next, for each of the three primary categories of ChatGPT-related projects (see Section~\ref{subsec:RQ1}), e.g., $c_j$, we measured the \textit{popularity} of an issue topic $z_k$ within $c_j$ as the proportion of issues whose dominant topic is $z_k$, i.e.,
\begin{equation}
\label{eq:popularity}
    popularity(z_k, c_j) = \frac{|Issues(c_j, z_k)|}{|Issues(c_j)|}
\end{equation}
where $Issues(c_j)$ represents the set of all issues from the projects within $c_j$; and $Issues(c_j, z_k)$ represents the subset of issues whose dominant topic is $z_k$, i.e., $\{d_i|d_i\in Issues(c_j) \land dominant\_topic(d_i) = z_k\}$.

\subsection{Difficulty Analysis of Issue Topics within Each Project Category} \label{subsec:difficulty analysis}
Intuitively, if an issue has received much attention but it has not been closed yet, then the issue should be probably difficult to address. The attention of an issue can be reflected by the number of comments and the number of participants associated with the issue. Based on these ideas, for each primary category of ChatGPT-related projects, $c_j$, we investigated the difficulty in addressing the issues of each topic $z_k$ within $c_j$ by measuring the average attention and closed rate of all issues belonging to $z_k$, as described as follows.

\textbf{Average number of comments.} Within $c_j$, we calculated the average number of comments for $z_k$ as the ratio of the total number of comments on the issues belonging to $z_k$ to the number of all issues belonging to $z_k$, i.e., 
\begin{equation}
\label{eq:avg comment}
    avg\_num_{comments}(c_j,z_k) = \frac{\sum_{d_i\in Issues(c_j,z_k)} num_{comments}(d_i)}{|Issues(c_j,z_k)|}
\end{equation}
where $num_{comments}(d_i)$ represents the number of comments on the issue $d_i$.

\textbf{Average number of participants.} Within $c_j$, we calculated the average number of participants for $z_k$ as the ratio of the total number of participants who commented on the issues belonging to $z_k$ to the number of all issues belonging to $z_k$, i.e.,
\begin{equation}
\label{eq:avg participants}
    avg\_num_{participants}(c_j, z_k) = \frac{\sum_{d_i\in Issues(c_j,z_k)} num_{participants}(d_i)}{|Issues(c_j,z_k)|}
\end{equation}
where $num_{participants}(d_i)$ represents the number of participants who commented on the issue $d_i$.




\textbf{Closing rate.} Within $c_j$, we calculated the closing rate for $z_k$ as the ratio of the total number of closed issues belonging to $z_k$ to the number of all issues belonging to $z_k$, i.e.,
\begin{equation}
\label{eq:closing rate}
    avg\_rate_{closed}(c_j,z_k) = \frac{\sum_{d_i\in Issues(c_j,z_k)} I(state(d_i)== closed)}{|Issues(c_j,z_k)|}
\end{equation}
where $state(d_i)$ represents the state of the issue $d_i$; and $I(condition)$ is an indicator function whose value is 1 if the $condition$ is true, otherwise 0.

\subsection{Evolution Analysis of Issue Topics within Each Project Category} \label{subsec:trend analysis}

The hotspots of issues will change during the progress of a project. We explored the evolution of the issue topics within each of the three primary project categories. For each issue topic, $z_k$, within a specific project category, $c_j$, we first defined and calculated the monthly variation of the impact of $z_k$ based on prior work~\cite{Wan2021} as 
\begin{equation}
\label{eq:impact}
    impact(c_j, z_k, m) = \frac{1}{|Issues(c_j,z_k,m)|}\sum_{d_i \in Issues(c_j,z_k,m)} \theta(d_i, z_k)
\end{equation}
where $Issues(c_j,z_k,m)$ represents the issues belonging to $z_k$ that are raised for the projects in $c_j$ in a specific month, $m$. 

Afterwards, we performed the Mann-Kendall trend test (MK test) to determine the evolution trend in the impact of each issue topic at a significance level of 0.05 using the \verb|pymannkendall| package\footnote{\url{https://pypi.org/project/pymannkendall/}}. The MK test is a non-parametric statistical test method for assessing trend changes in time series data. Additionally, we calculated the Theil-Sen slope~\cite{Shourov2019} to quantify the magnitude of the monotonic trend, which is often used in conjunction with the MK test. Based on the results, we identified the issue topics that exhibit increasing or decreasing trends and inferred their causes.










\section{Results} \label{sec:results}



\subsection{RQ1. What categories of ChatGPT-related projects have been developed?} \label{subsec:RQ1}

According to the categorization method described in Section~\ref{subsec:project categorization}, we identified four categories of ChatGPT-related projects: \textit{ChatGPT Implementation \& Training}, \textit{ChatGPT Application}, \textit{ChatGPT Improvement \& Extension}, \textit{Other ChatGPT-related Project}, which are explained as follows. 

{\bfseries ChatGPT Implementation \& Training}. The projects in this category focus on studying the underlying model of ChatGPT and implementing it through other large language models. Additionally, they explore ways to improve the training process of ChatGPT, including optimizing training methods for model parameters, fine-tuning the model to adapt to different downstream tasks, and constructing datasets for model training. 
For example, the project `PhoebusSi/Alpaca-CoT'\footnote{\url{https://github.com/PhoebusSi/Alpaca-CoT}} builds a fine-tuning platform which is easy to get started and use.

{\bfseries ChatGPT Application}. The projects in this category aim to unlock the potential capabilities of ChatGPT by leveraging it to solve various tasks, 
e.g., text translation and code generation. Moreover, 
calling ChatGPT APIs to develop conversational robots has also received widespread attention. 
For example, the project `gragland/chatgpt-chrome-extension'\footnote{\url{https://github.com/gragland/chatgpt-chrome-extension}} develops a tool for integrating ChatGPT into text boxes on the Internet.

{\bfseries ChatGPT Improvement \& Extension}. The projects in this category improve ChatGPT from different aspects, including the calling of ChatGPT APIs and the functional expansion and performance enhancement of ChatGPT. Specifically, developers encapsulate the ChatGPT APIs for easier application calls, assist users in resolving various network issues that may arise when accessing ChatGPT, provide prompt templates to optimize ChatGPT’s replies, and use network search results to enhance the accuracy of results generated by ChatGPT. 
For example, the project `Yidadaa/ChatGPT-Next-Web'\footnote{\url{https://github.com/Yidadaa/ChatGPT-Next-Web}} develops a cross-platform user interface for ChatGPT.

{\bfseries Other ChatGPT-related Project}. This category contains the other ChatGPT-related projects that cannot be classified into the three categories above, e.g., 
a collection of ChatGPT mirror websites, survey reports about ChatGPT, etc. For example, the project `LiLittleCat/awesome-free-chatgpt'\footnote{\url{https://github.com/LiLittleCat/awesome-free-chatgpt}} collects a list of free ChatGPT mirror websites.

\begin{table}
  \caption{The numbers of ChatGPT-related projects within four categories}
  \label{tab:number of projects}
  \begin{tabular}{lc}
    \toprule
    Category & \#Projects\\
    \midrule
    ChatGPT Implementation \& Training & 8\\
    ChatGPT Application & 55\\
    ChatGPT Improvement \& Extension & 55\\
    Other ChatGPT-related Project & 18\\
  \bottomrule
\end{tabular}
\end{table}

\begin{figure}
  \centering
  \includegraphics[width=.75\linewidth]{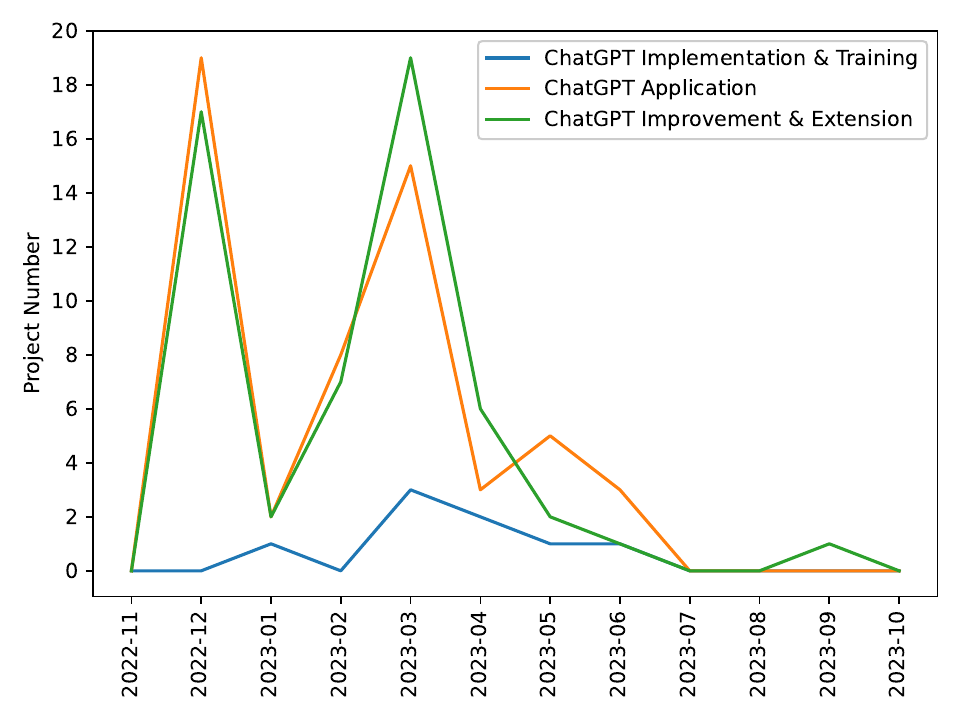}
  \caption{The monthly increased numbers of projects within three primary categories}
  \label{fig:project num change}
\end{figure}

Table~\ref{tab:number of projects} presents the number of projects assigned to the four categories. In addition, we counted the number of projects growing in every month for the three primary categories: \textit{ChatGPT Implementation \& Training}, \textit{ChatGPT Application}, \textit{ChatGPT Improvement \& Extension}, as depicted in Fig.~\ref{fig:project num change}. As can be seen, \textbf{the monthly increase in the number of projects within the \textit{ChatGPT Application} and \textit{ChatGPT Improvement \& Extension} categories peaked in December 2022 and March 2023.} 
This phenomenon is closely related to the evolution of ChatGPT. Specifically, after OpenAI launched the first version of ChatGPT built on the GPT-3.5 model on November 30, 2022, there was a significant increase in the number of ChatGPT-related projects on GitHub. The first projects in the two categories were created on December 2, 2022 and December 1, 2022, respectively, followed by a rapid growth in that month. 
Subsequently, OpenAI launched the multi-modal model GPT-4 in March 2023, which significantly enhanced the capabilities of GPT-3.5 and further stimulated the enthusiasm of researchers and developers. For example, in the issue \#471\footnote{\url{https://github.com/transitive-bullshit/chatgpt-api/issues/471}} (created on March 15, 2023) of the project `transitive-bullshit/chatgpt-api' (created on December 3, 2022), the user hoped that GPT-4 could be incorporated into the project. The project `enricoros/big-agi'\footnote{\url{https://github.com/enricoros/big-agi}} created on March 19, 2023 is a personal AI application powered by GPT-4. As a result, the growing number of projects in the \textit{ChatGPT Application} and \textit{ChatGPT Improvement \& Extension} categories reached the second peak in March 2023. 
Later, OpenAI launched the iOS version of ChatGPT in May 2023, which also promoted the secondary development of ChatGPT to a certain extent. For example, in the issue \#142\footnote{\url{https://github.com/sunner/ChatALL/issues/142}} (created on May 26, 2023) of the project `sunner/ChatALL', the user hoped that developers could add the iOS client models of ChatGPT to their project. {\bfseries Therefore, in May 2023, the number of projects in the \textit{ChatGPT Application} category saw another increase.} 

Moreover, when ChatGPT was initially released, most developers focused on improving the use of ChatGPT or exploring ways to conduct secondary development based on ChatGPT. As time progressed, an increasing number of open-source large language models began to emerge, making customized implementation and training of ChatGPT possible. For example, the project `mymusise/ChatGLM-Tuning'\footnote{\url{https://github.com/mymusise/ChatGLM-Tuning}} (created on March 16, 2023) used the Tsinghua's ChatGLM-6B model as an affordable solution for ChatGPT implementation. {\bfseries Thus, a peak in the growing number of projects in the \textit{ChatGPT Implementation \& Training} category occurred in March 2023.}

\begin{center}
    \fcolorbox{black}{gray!20}{\parbox{\textwidth}{
    {\bfseries Summary:} We classified 200 popular open-source projects searched using the keyword `ChatGPT' into five categories. The three primary categories of ChatGPT-related projects are \textit{ChatGPT Implementation \& Training}, \textit{ChatGPT Application}, \textit{ChatGPT Improvement \& Extension}. 
    The monthly increased numbers of projects assigned to \textit{ChatGPT Application} and \textit{ChatGPT Improvement \& Extension} are closely associated with the development of ChatGPT. Moreover, the number of projects in \textit{ChatGPT Implementation \& Training} relates to the development of open-source large language models.}}
\end{center}

\subsection{RQ2. What are the topics of issues discussed in ChatGPT-related projects?} \label{subsec:RQ2}

\begin{figure}
  \centering
  \includegraphics[width=\linewidth]{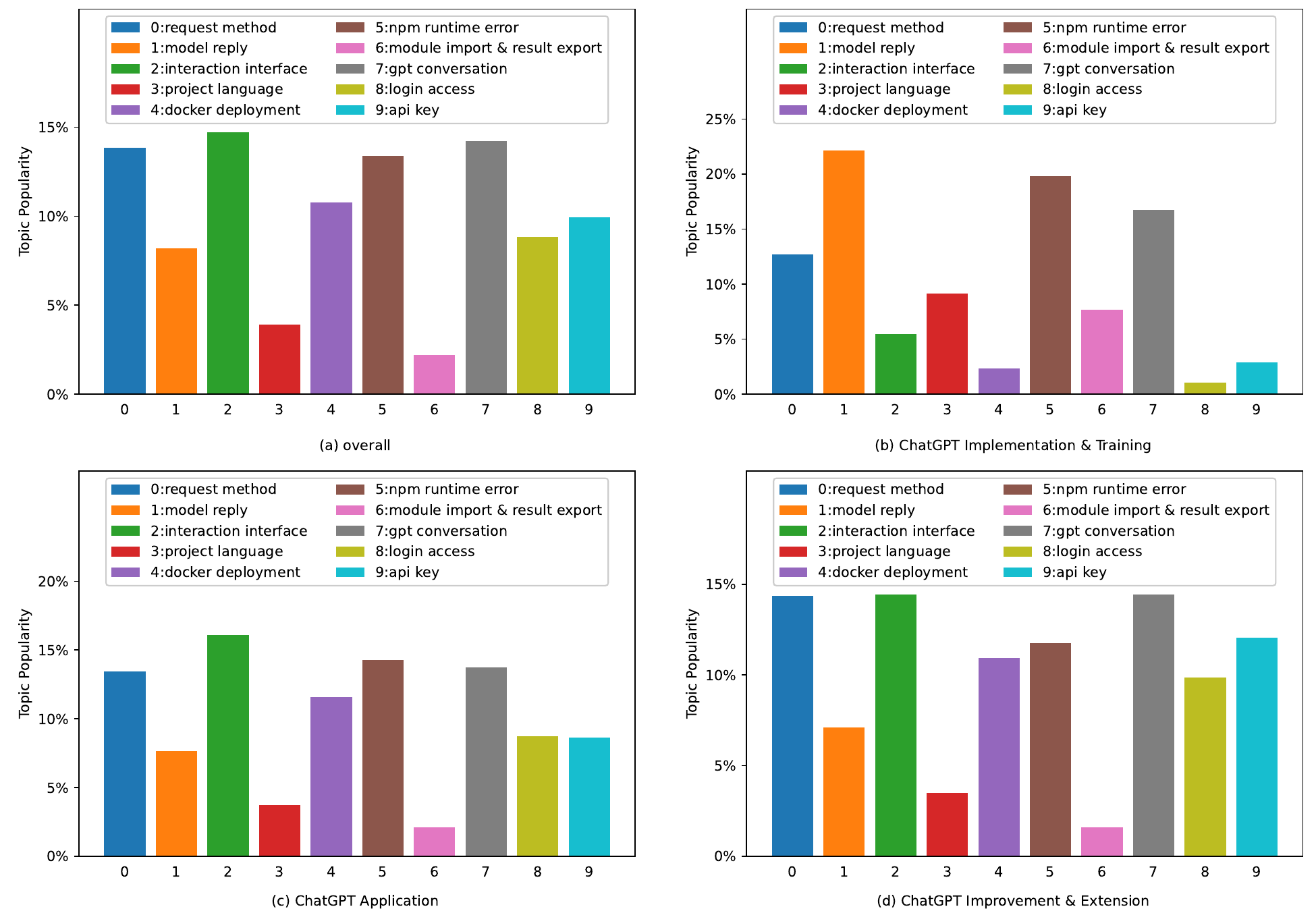}
  \caption{The popularity of issue topics in the overall dataset and within the three primary project categories}
  \label{fig:topic popularity}
\end{figure}

As described in Section~\ref{subsec:train LDA}, in order to understand the issues reported about ChatGPT-related projects, we applied the LDA topic model to the dataset of 23,609 issues collected from the projects in the three primary categories identified in RQ1. Table~\ref{tab:topics of issue texts} presents the ten discovered issue topics and their top 10 keywords. Moreover, we calculated the popularity of each topic in the overall dataset and the three primary project categories using Eq.~(\ref{eq:popularity}), as shown in Fig.~\ref{fig:topic popularity}. 

\begin{table}
  \caption{The issue topics with their top ten keywords}
  \label{tab:topics of issue texts}
  \begin{tabular}{ll}
    \toprule
    Topic name & Top ten keywords\\
    \midrule
    request method & request featur function support chatgpt add user api share enhanc\\
    model reply & model answer amd output inform prompt interfac time paramet data\\
    interaction interface & bug click open button window time display file input work\\
    project language & project languag ai model version chines support pictur chrome work\\
    docker deployment & error chatgpt docker deploy api messag bug send openai web\\
    npm runtime error & error run npm instal version work file warn build code\\
    module import \& result export & import export line administr client arm imag slow error speed\\
    gpt conversation & gpt chat convers prompt add model messag user set make\\
    login access & error server login access fail proxi connect agent password http\\
    api key & api token key dialogu word alloc access mode modifi prompt\\
  \bottomrule
\end{tabular}
\end{table}


Fig.~\ref{fig:topic popularity}(a) shows the popularity of issue topics in the overall dataset. We can see that the popularity of the four topics, namely \textit{request method}, \textit{interaction interface}, \textit{npm runtime error}, and \textit{gpt conversation}, is greater than 13\%. The popularity of the other four topics, i.e., \textit{model reply}, \textit{docker deployment}, \textit{login access}, and \textit{api key}, is greater than 8\%. The popularity of \textit{project language} and \textit{module import \& result export} is the lowest, with only 3.9\% and 2.2\%, respectively. Due to the fact that the \textit{ChatGPT Application} and \textit{ChatGPT Improvement \& Extension} categories account for a large proportion of ChatGPT-related projects, the issues are largely relevant to the improvement and secondary development of ChatGPT. For example, there is significant focus on improving the interactive interface of ChatGPT, the methods of sending requests to ChatGPT, and communicating with ChatGPT to obtain the conversation results, etc. 

Fig.~\ref{fig:topic popularity}(b) shows the popularity of each topic within the \textit{ChatGPT Implementation \& Training} category, which notably differs from the topic popularity in the overall dataset. Specifically, \textbf{the popularity of the \textit{model reply}, \textit{project language}, \textit{npm runtime error}, and \textit{module import \& result export} topics is obviously higher than the others.} In particular, \textit{model reply} has the highest popularity of 22.1\%. Conversely, the popularity of \textit{interaction interface}, \textit{docker deployment}, \textit{login access}, and \textit{api key}, is much lower than that shown in Fig.~\ref{fig:topic popularity}(a). This is because the projects in the \textit{ChatGPT Implementation \& Training} category primarily focus on implementing and training ChatGPT-like models. Consequently, they pay more attentions to the replies generated by the model and the storage of output results. For example, when LoRA technology was applied to fine-tune the model, a user raised an issue about the responses being too short\footnote{\url{https://github.com/PhoebusSi/Alpaca-CoT/issues/111}}. Developers can fine-tune or retrain the model based on the exported results to make it closer to ChatGPT. 


Fig.~\ref{fig:topic popularity}(c) shows the popularity of issue topics within the \textit{ChatGPT Application} category. \textbf{Compared to the results shown in Fig.~\ref{fig:topic popularity}(a), in spite of \textit{interaction interface} (whose popularity is higher) and \textit{api key} (whose popularity is lower), there is no significant difference in the popularity of the other topics.} The projects in this category typically use ChatGPT as the underlying model or call ChatGPT APIs for development, thereby integrating the capabilities of ChatGPT into their application tools. As an application tool, users pay more attention to the interaction interface and may also propose new interaction requirements. For example, in the issue \#1006\footnote{\url{https://github.com/binary-husky/gpt_academic/issues/1006}} of the project `binary-husky/gpt\_academic', the user hoped to add a small button next to the submit button to decide whether to carry historical information during conversations. Although the projects in this category frequently call ChatGPT APIs to utilize its functionalities, which might lead to more issues about \textit{api key}, the official documentation from OpenAI provides detailed instructions on how to apply for and use the ChatGPT APIs with examples. Therefore, users can solve most of \textit{api key} related issues, which may explain the low popularity of the \textit{api key} topic.


Fig.~\ref{fig:topic popularity}(d) shows the popularity of issue topics within the \textit{ChatGPT Improvement \& Extension} category. \textbf{Compared to the results shown in Fig.~\ref{fig:topic popularity}(a), the popularity of \textit{login access} and \textit{api key} is higher, while the popularity of \textit{model reply} and \textit{npm runtime error} is lower.} 
This is because the goal of \textit{ChatGPT Improvement \& Extension} is to facilitate the use of ChatGPT. Therefore, providing convenient login and access methods, as well as improving the methods for calling ChatGPT APIs, are important in the projects of this category. For example, in the issue \#363\footnote{\url{https://github.com/Yidadaa/ChatGPT-Next-Web/issues/363}} and \#1022\footnote{\url{https://github.com/Yidadaa/ChatGPT-Next-Web/issues/1022}} of the project `Yidadaa/ChatGPT-Next-Web', users expressed a desire to add multi-api key polling functionality. 

\begin{center}
    \fcolorbox{black}{gray!20}{\parbox{\textwidth}{
    {\bfseries Summary:} We discovered ten topics, e.g., \emph{request method} and \emph{model reply}, from the issues of the three primary categories of ChatGPT-related projects. There are significant differences in the popularity of topics within and across different categories. For example, the \textit{model reply} and \textit{interaction interface} topics respectively receive the highest popularity in the \textit{ChatGPT Implementation \& Training} and \textit{ChatGPT Application} categories. The popularity of \textit{api key} in the \textit{ChatGPT Improvement \& Extension} category is much higher than those in the other two categories.
    }}
\end{center}

\subsection{RQ3. How difficult is it to address the issues of each topic within different project categories?} \label{subsec:RQ3}

We defined and measured three metrics, i.e., average number of comments ($\mathrm{avg\_num}_{comments}$), average number of participants ($\mathrm{avg\_num}_{participants}$), and closing rate ($\mathrm{avg\_rate}_{closed}$), to investigate the difficulty in addressing the issues of each topic within the three primary ChatGPT-related project categories, as described in Section~\ref{subsec:difficulty analysis}. These metrics reflect the attention and solution status of the issue topics. Generally, a topic with higher attention and a lower solution rate probably indicates that the issues of the topic are more difficult to address.

Table~\ref{tab:attention for implementation} presents the metric results measured for each issue topic within the \textit{ChatGPT Implementation \& Training} category. Particularly, the \textit{model reply} topic has the highest popularity (see Fig.~\ref{fig:topic popularity}(b)) and also leads in both average number of comments and average number of participants, however its closing rate is the lowest. This result means that the issues related to this topic are difficult to address since they received high attention, but only a low proportion of them have been resolved. We believe this phenomenon occurs because different underlying models and datasets could be used for training, which leads to unexpected results in the implementation of models similar to ChatGPT. For example, in the issue \#617\footnote{\url{https://github.com/OptimalScale/LMFlow/issues/617}} of the project `OptimalScale/LMFlow', the user used a LLM (`Chinese-Llama2-7b') to test the fine-tuning ability of the tool. However, this model is different from the underlying model of ChatGPT, which results in output errors. 
Additionally, the training process requires multiple rounds of fine-tuning which depends on the dataset and the model response. This not only requires a considerable amount of time, but may also yield different results during every training session, which increases the difficulty in solving related issues.

\begin{table}
  \caption{Attention and solution status of each issue topic within the \emph{ChatGPT Implementation \& Training} category}
  \label{tab:attention for implementation}
  \begin{threeparttable}
  \begin{tabular}{lcccc}
    \toprule
    Topic name & \#Issues & $\mathrm{avg\_num}_{comments}$ & $\mathrm{avg\_num}_{participants}$ & $\mathrm{avg\_rate}_{closed}$ \\
    \midrule
    request method                 & 168         & 1.86                      & 1.33                        & 68.5\%                          \\
    model reply                    & 292         & 3.37                      & 2.18                        & 57.2\%                          \\
    interaction interface          & 72          & 2.86                      & 1.81                        & 72.2\%                          \\
    project language               & 121         & 2.4                       & 1.68                        & 70.2\%                          \\
    docker deployment              & 31          & 2.94                      & 1.77                        & 80.6\%                          \\
    npm runtime error              & 262         & 3.05                      & 1.88                        & 81.7\%                          \\
    module import \& result export & 101         & 3.68                      & 2.13                        & 77.2\%                          \\
    gpt conversation               & 221         & 2.14                      & 1.34                        & 75.6\%                          \\
    login access                   & 14          & 3.29                      & 2.14                        & 71.4\%                          \\
    api key                        & 38          & 2.53                      & 1.68                        & 63.2\%                          \\
    \bottomrule
    \end{tabular}
    \end{threeparttable}
\end{table}

\begin{table}
  \caption{Attention and solution status of each issue topic within the \emph{ChatGPT Application} category}
  \label{tab:attention for application}
  \begin{threeparttable}
  \begin{tabular}{lcccc}
    \toprule
   Topic name & \#Issues & $\mathrm{avg\_num}_{comments}$ & $\mathrm{avg\_num}_{participants}$ & $\mathrm{avg\_rate}_{closed}$ \\
    \midrule
    request method                 & 1,497        & 2.28                      & 1.47                        & 66.3\%                          \\
    model reply                    & 854         & 3                         & 1.83                        & 75.5\%                          \\
    interaction interface          & 1,795        & 2.76                      & 1.7                         & 72.7\%                          \\
    project language               & 414         & 2.24                      & 1.5                         & 67.4\%                          \\
    docker deployment              & 1,292        & 3.57                      & 2.26                        & 81.7\%                          \\
    npm runtime error              & 1,594        & 2.96                      & 1.85                        & 77.7\%                          \\
    module import \& result export & 238         & 2.64                      & 1.62                        & 72.3\%                          \\
    gpt conversation               & 1,530        & 2.45                      & 1.5                         & 76.9\%                          \\
    login access                   & 976         & 3.76                      & 2.31                        & 78.3\%                          \\
    api key                        & 965         & 2.88                      & 1.97                        & 74.9\%                          \\
    \bottomrule
    \end{tabular}
    \end{threeparttable}
\end{table}

\begin{table}
  \caption{Attention and solution status of each issue topic within the \emph{ChatGPT Improvement \& Extension} category}
  \label{tab:attention for improvement}
  \begin{threeparttable}
  \begin{tabular}{lcccc}
    \toprule
    Topic name & \#Issues & $\mathrm{avg\_num}_{comments}$ & $\mathrm{avg\_num}_{participants}$ & $\mathrm{avg\_rate}_{closed}$ \\
    \midrule
    request method                 & 1,600        & 2.86                      & 1.87                        & 74.8\%                          \\
    model reply                    & 790         & 3.25                      & 2.16                        & 92.2\%                          \\
    interaction interface          & 1,608        & 3.08                      & 1.95                        & 75.7\%                          \\
    project language               & 388         & 2.65                      & 1.77                        & 78.4\%                          \\
    docker deployment              & 1,218        & 4.25                      & 2.5                         & 84.9\%                          \\
    npm runtime error              & 1,308        & 3.31                      & 2.04                        & 76.4\%                          \\
    module import \& result export & 178         & 2.98                      & 1.97                        & 83.1\%                          \\
    gpt conversation               & 1,606        & 2.69                      & 1.74                        & 61.9\%                          \\
    login access                   & 1,096        & 4.1                       & 2.61                        & 88.0\%                          \\
    api key                        & 1,342        & 3.35                      & 2.2                         & 88.8\%                          \\
    \bottomrule
    \end{tabular}
    \end{threeparttable}
\end{table}

Table~\ref{tab:attention for application} presents the metric results of each issue topic within the \textit{ChatGPT Application} category. The \textit{interaction interface}, the most popular topic in this category (see Fig.~\ref{fig:topic popularity}(c)), ranks about halfway among all topics in terms of average number of comments and average number of participants, indicating that the issues related to this topic received relatively high attention. Additionally, the closing rate of \textit{interaction interface} reaches 72.7\%, showing that most of the issues raised by users have been addressed. This could be explained by the reason that the projects in the \textit{ChatGPT Application} category are primarily application tools. Addressing the issues related to the interaction interface is beneficial for the promotion and usability of these tools.

In contrast, another popular topic \textit{request method} has the lowest closing rate. This may be because the projects often need to send various types of requests. For example, the project `sunner/ChatALL'\footnote{\url{https://github.com/sunner/ChatALL}} is able to communicate simultaneously with many large language models, such as ChatGPT, Bing Chat, ChatGLM, and MOSS, to find the best answer. However, different types of requests have different parameter settings and rules, which leads to many unpredictable issues. For example, in the issue \#301\footnote{\url{https://github.com/sunner/ChatALL/issues/301}} of the project `sunner/ChatALL’, the user found that `claude-v1’ model disconnects every ten minutes and requires a manual refresh to solve this issue. Therefore, the issues related to this topic are often difficult to address.

Table~\ref{tab:attention for improvement} presents the metric results of each issue topic within the \textit{ChatGPT Improvement \& Extension} category. The average number of comments and participants of each topic are higher than those in the other two categories, indicating that these topics received higher attention. Additionally, except for \textit{gpt conversation}, the closing rates of the other topics are all close to or above 75\% and significantly exceed those of the corresponding topics in the other two categories. This shows that most issues in this category have been resolved. This could be explained by the reason that the developers in this category need to have a good understanding of ChatGPT and thus can quickly address the issues.

Moreover, the popular topic \textit{gpt conversation} (see Fig.~\ref{fig:topic popularity}(d)) has the lowest closing rate, i.e., 61.9\%. This is because in the \textit{ChatGPT Improvement \& Extension} category, most issues of this topic concern improving the conversation process. Users often request new features to enhance their experience. For example, users may wish to have multiple sessions simultaneously rather than following the same pattern as on the ChatGPT official website\footnote{\url{https://github.com/mckaywrigley/chatbot-ui/issues/107}}. Implementing new features could be a challenge for developers, which increases the difficulty in addressing these issues.

\begin{center}
    \fcolorbox{black}{gray!20}{\parbox{\textwidth}{
    {\bfseries Summary:} We measured the difficulty in addressing the issues of the ten topics discovered previously and found notable differences across the three primary categories of ChatGPT-related projects. For example, in the \textit{ChatGPT Implementation \& Training} category, although the \textit{model reply} topic received high attention, its closing rate is the lowest, indicating that the issues of this topic are difficult to solve. In the \textit{ChatGPT Application} category, the existence of multiple different request types complicates the solving of issues related to the \textit{request method} topic. In the \textit{ChatGPT Improvement \& Extension} category, except for the \textit{gpt conversation} topic, most issues of the other topics are well addressed.
    }}
\end{center}

\subsection{RQ4. How do the issue topics evolve over time within different project categories?} \label{subsec:RQ4}

As described in Section~\ref{subsec:trend analysis}, we first calculated the monthly changes in the impact of each issue topic within the three primary ChatGPT-related project categories. Then, we used the MK test to determine the evolution trend of each topic. 

Table~\ref{tab:mk test implementation} presents the evolution trend of the impact of each issue topic within the \textit{ChatGPT Implementation \& Training} category. \textbf{The impact of \textit{model reply} shows a decreasing trend, while the impact of \textit{npm runtime error} and \textit{gpt conversation} exhibits an increasing trend.} The impact of the other topics does not show significant trends. 
This is because the \textit{model reply} topic is related to the training phase of a ChatGPT-like model. In the early stages of a project, there is a significant gap between the open-source model used to implement ChatGPT and the actual ChatGPT model. Both developers and users need to train and fine-tune the model based on its responses. As the process improves, the model gradually becomes similar to the ChatGPT model. Thus, the impact of \textit{model reply} gradually decreases; thereafter users shift their attention to the configuration and use of the model, which leads to more issues about \textit{npm runtime error} and \textit{gpt conversation}. For example, users can install the JavaScript packages of the project through npm commands\footnote{\url{https://github.com/embedchain/embedchain/issues/122}}. In the issue \#35\footnote{\url{https://github.com/Instruction-Tuning-with-GPT-4/GPT-4-LLM/issues/35}} of the project `Instruction-Tuning-with-GPT-4/GPT-4-LLM', the user reported that there exist a considerable number of low-quality dialogue samples in `unnatural\_instruction\_gpt4\_data.json'. Therefore, the impact of these two issue topics increases accordingly. Moreover, since the other topics are almost unrelated to the early training and later use of the model, there are no significant trends in their impact.

\begin{table}
  \caption{Evolution trend of each issue topic within the \emph{ChatGPT Implementation \& Training} category}
  \label{tab:mk test implementation}
  \begin{tabular}{lccc}
    \toprule
    Topic name                      & Evolution trend      & p-value                     & Sen's slope \\
    \midrule
    request method                  & no trend   & 0.063486531                 & 0.020925277                     \\
    model reply                     & decreasing & 0.004434008                 & -0.046970927                    \\
    interaction interface           & no trend   & 0.265510389                 & 0.00461322                      \\
    project language                & no trend   & 0.10776229                  & -0.006612922                    \\
    docker deployment               & no trend   & 0.901538627                 & 7.2633E-05                      \\
    npm runtime error               & increasing & 0.035447893                 & 0.011221673                     \\
    module import \& result export & no trend   & 0.063486531                 & -0.009560359                    \\
    gpt conversation                & increasing & 0.009374768                 & 0.040220779                     \\
    login access                    & no trend   & 0.10776229                  & -0.001600849                    \\
    api key                         & no trend   & 0.901538627                 & -0.001691792\\
    \bottomrule
\end{tabular}
\end{table}

\begin{table}
  \caption{Evolution trend of each issue topic within the \emph{ChatGPT Application} category}
  \label{tab:mk test application}
  \begin{tabular}{lccc}
    \toprule
    Topic name                      & Evolution trend      & p-value & Sen's slope \\
    \midrule
    request method                  & increasing & 0.003093449                 & 0.005350561                     \\
    model reply                     & no trend   & 0.086768173                 & 0.001541284                     \\
    interaction interface           & increasing & 0.003093449                 & 0.010677818                     \\
    project language                & no trend   & 0.119470987                 & 0.001846343                     \\
    docker deployment               & decreasing & 0.000613905                 & -0.01015641                     \\
    npm runtime error               & no trend   & 0.275757848                 & -0.003741957                    \\
    module import \& result export & no trend   & 0.876269664                 & 5.49696E-05                     \\
    gpt conversation                & increasing & 0.00506931                  & 0.008503926                     \\
    login access                    & decreasing & 0.001845721                 & -0.006764936                    \\
    api key                         & decreasing & 0.012731364                 & -0.004841313\\
    \bottomrule
\end{tabular}
\end{table}

\begin{table}
  \caption{Evolution trend of each issue topic within the \emph{ChatGPT Improvement \& Extension} category}
  \label{tab:mk test improvement}
  \begin{tabular}{lccc}
    \toprule
    Topic name                      & Evolution trend      & p-value & Sen's slope \\
    \midrule
    request method                  & no trend   & 0.161124949                 & 0.000929693                     \\
    model reply                     & no trend   & 0.533416513                 & 0.002768807                     \\
    interaction interface           & no trend   & 0.161124949                 & -0.002702788                    \\
    project language                & increasing & 0.042960146                 & 0.001112902                     \\
    docker deployment               & no trend   & 0.640428787                 & 0.000223462                     \\
    npm runtime error               & decreasing & 0.042960146                 & -0.008559295                    \\
    module import \& result export & no trend   & 0.876269664                 & 0.000103658                     \\
    gpt conversation                & no trend   & 0.350201389                 & -0.002809677                    \\
    login access                    & no trend   & 0.350201389                 & 0.001377567                     \\
    api key                         & no trend   & 0.086768173                 & 0.008431346\\
    \bottomrule
\end{tabular}
\end{table}

Table~\ref{tab:mk test application} presents the evolution trend of the impact of each issue topic within the \textit{ChatGPT Application} category. \textbf{As time goes by, the impact of \textit{request method}, \textit{interaction interface}, and \textit{gpt conversation} shows an increasing trend. Conversely, the impact of \textit{docker deployment}, \textit{login access}, and \textit{api key} shows a decreasing trend.} The impact of the other topics does not show significant trends. This is because the projects in the \textit{ChatGPT Application} category aim to apply ChatGPT in various tasks. After developers complete the development of a project, users usually need to install and deploy the project based on its README file and then configure the ChatGPT APIs to login and use the project. Therefore, in the early stages of a project, users mainly focus on issues related to deployment, configuration, and login access. For example, the issues \#127\footnote{\url{https://github.com/fuergaosi233/wechat-chatgpt/issues/127}}, \#130\footnote{\url{https://github.com/fuergaosi233/wechat-chatgpt/issues/130}}, and \#250\footnote{\url{https://github.com/fuergaosi233/wechat-chatgpt/issues/250}} of the project `fuergaosi233/wechat-chatgpt' are all related to the docker deployment. As the project improves and developers provide answers to these initial issues, they are no longer a hindrance for users. Thus, users start to pay more attention to the interaction interface and the quality of conversation. This shift leads to a decreasing trend in the impact of \textit{docker deployment}, \textit{login access}, and \textit{api key}, and an increasing trend in the impact of \textit{request method}, \textit{interaction interface}, and \textit{gpt conversation}.

Table~\ref{tab:mk test improvement} presents the evolution trend of the impact of each issue topic within the \textit{ChatGPT Improvement \& Extension} category. \textbf{Over time, the impact of \textit{project language} shows an increasing trend, while the impact of \textit{npm runtime error} shows a decreasing trend.} The impact of the other topics does not show significant trends. This is because the projects in the \textit{ChatGPT Improvement \& Extension} category focus on enhancing various aspects of ChatGPT, such as usage, functionality extension, and performance improvement. Consequently, these projects often introduce additional software packages to facilitate the use of ChatGPT and extend its functionalities. However, software packages in the npm module repository may undergo updates and rollbacks, leading to incompatibility and crashes. Therefore, the issues related to \textit{npm runtime error} often occur in the early stages of a project. For example, a user encountered an error `the module could not be found' when executing the command `npm run dev'\footnote{\url{https://github.com/pashpashpash/vault-ai/issues/47}}. But with the improvement of the project and the introduction of more stable npm software packages, the issues gradually decrease. Moreover, ChatGPT interacts with humans through dialogue, which requires accepting various types of language and text inputs. Thus, the impact of \textit{project language} is increasing. For example, a user expressed a desire for the project `ztjhz/BetterChatGPT' to provide a LaTeX rendering switch to render inline formulas to improve the readability\footnote{\url{https://github.com/ztjhz/BetterChatGPT/issues/251}}. As for the topics \textit{request method}, \textit{interaction interface}, \textit{gpt conversation}, \textit{login access}, and \textit{api key}, which are important aspects of ChatGPT improvement, their impact does not show significant trends.

\begin{center}
    \fcolorbox{black}{gray!20}{\parbox{\textwidth}{
    {\bfseries Summary:} Over time, the impact of the ten issue topics shows different evolution trends within the three primary categories of ChatGPT-related projects. For example, in the \textit{ChatGPT Implementation \& Training} category, the impact of \textit{model reply} shows a decreasing trend, while the impact of \textit{npm runtime error} and \textit{gpt conversation} shows an increasing trend. In the \textit{ChatGPT Application} category, the impact of \textit{interaction interface} and \textit{docker deployment} shows an increasing trend and a decreasing trend, respectively. In the \textit{ChatGPT Improvement \& Extension} category, the impact of \textit{project language} and \textit{npm runtime error} shows an increasing trend and a decreasing trend, respectively.
    }}
\end{center}

\section{Discussion} \label{sec:discussion}

In this section, we first discuss the implications for the management platform and developers of ChatGPT-related projects based on our research results and then discuss the threats to the validity of this work.

\subsection{Implications}\label{subsec:implications}


{\bfseries Implications for project management platforms}. The categories of ChatGPT-related projects can facilitate the organization and exploration of the projects on GitHub or other platforms (e.g., Gitee). For example, when developers tag a project with `chatgpt', the platform can suggest a more fine-grained tag, e.g., `ChatGPT Implementation \& Training' or `ChatGPT Application'. This can not only optimize the management of ChatGPT-related projects but also refine the search results for users, thereby enhancing their retrieval efficiency.

{\bfseries Implications for developers}. There are several implications for developers to develop, improve, and promote their ChatGPT-related projects. 
(i) {\bfseries Focusing on the application and improvement of ChatGPT.} On the GitHub platform, most ChatGPT-related projects are associated with developing ChatGPT-based applications and improving the use of ChatGPT. These projects enable practitioners in various fields to benefit from ChatGPT, which has high economic and practical value and is worthy of further research.  
(ii) {\bfseries Staying updated with ChatGPT developments.} The development of ChatGPT is crucial for developing and improving ChatGPT-related projects. For example, after the release of GPT-4, developers should promptly integrate it into their projects promptly and continuously stay aligned with the development of ChatGPT, which can quickly meet user requirements and enhance project visibility. 
(iii) {\bfseries Attention should vary across different categories of ChatGPT-related projects.} The popularity and difficulty of issue topics are different within different project categories. Developers should pay more attention to the popular and difficult issues of projects in each category. For example, in the \textit{ChatGPT Implementation \& Training} category, developers should focus more on issues related to \textit{model reply}. It is advisable to specify the dataset, underlying model, and parameter settings used during training in the README file to help users replicate the process. In the \textit{ChatGPT Application} category, developers should timely satisfy new requirements raised by users regarding \textit{interaction interface} and ensure that the descriptions of \textit{request method} in the README file are clear and complete. In the \textit{ChatGPT Improvement \& Extension} category, developers can enhance the experience of using ChatGPT by providing diverse interfaces and extending API calling methods. 
(iv) {\bfseries Adjusting development focus over time.} The development focus of a project should be adjusted over time since the impact of issue topics will evolve during the progress of the project. For example, in the \textit{ChatGPT Implementation \& Training} category, as the project progresses, the proportion of issues related to the \textit{gpt conversation} topic will increase. Developers need to focus on issues that contain keywords such as `gpt’, `chat’, and `conversation’. In the \textit{ChatGPT Application} category, the impact of \textit{interaction interface} topic will increase and thus developers need to prioritize addressing issues that contain keywords like `click’, `open’, and `button’.

\subsection{Threats to validity}

{\bfseries Threats to internal validity} relate to the errors in the implementation of our research methodology as well as the subjective bias of participants in the manual categorization of ChatGPT-related projects. To ensure the correctness of our research methodology, we double-checked the implementation code and execution of every step shown in Fig.~\ref{fig:research methodology}. For the categorization of the top 200 projects with the highest numbers of stars, the first and second co-authors of the paper performed the task using an open-card sorting approach in two phases, i.e., category determination and project classification, as described in Section~\ref{subsec:project categorization}. To minimize the subjective bias, both co-authors independently conducted each phase task and then resolved any inconsistencies through discussion to reach a common decision. 

{\bfseries Threats to external validity} relate to the generalizability of our results. In this work, we selected the top 200 open-source projects (searched using the keyword `ChatGPT') with the highest numbers of stars from GitHub, the world's largest project hosting platform, as our candidate dataset of ChatGPT-related projects. These projects cover a representative set of the most popular projects related to ChatGPT. Moreover, we collected a relatively large dataset of 23,609 issues from the 118 ChatGPT-related projects within three primary categories. These two popular and large datasets could help improve the generalizability of our results. 

\section{Related Work} \label{sec:related work}


\subsection{ChatGPT Studies}

There are many studies on the application of ChatGPT in different tasks, such as natural language processing (NLP), code analysis. Additionally, numerous studies have focused on the worries about ChatGPT.

{\bfseries NLP with ChatGPT.} The natural language understanding and generation capabilities of ChatGPT have received increasing attention from researchers. Firat~\cite{Firat2023HowChatGPTCanTransformAutodidacticExperiences} believed that ChatGPT would become a promising tool in the field of open education. Due to its ability to understand and respond to natural language, ChatGPT could provide personalized guidance and assistance to learners, thereby improving their ability to independently solve problems and increasing their self-learning motivation. Búadóttir et al.~\cite{Telma2023} developed a financial consulting tool that utilizes ChatGPT to answer questions. Users only need to submit their financial problems for consultation and resolution. Alessa and Al-Khalifa~\cite{Alessa2023} used the conversational interaction capability of ChatGPT to create conversation partners for the elderly, thereby helping the elderly reduce their loneliness and social isolation. Kozachek~\cite{Kozachek2023} explored the potential of GPT-3, GPT-3.5, and GPT-4 in describing future scenarios of human society and provided several artificial scene examples to improve the quality of outputs. Therefore, providing high-quality prompts for ChatGPT is crucial for optimizing its generated results. White et al.~\cite{Jules2023} examined several prompting patterns, such as improving requirements elicitation, improving code quality, improving system design, etc., and applied them to ChatGPT to help reduce and overcome common errors in software engineering tasks. However, possessing a certain amount of experience is essential for users to generate effective prompts. In order to reduce the time spent creating excellent prompts and improve the efficiency of using ChatGPT, Firat and Kuleli~\cite{Firat2023WhatifGPT4BecameAutonomous} developed Auto-GPT, which automates the utilization of GPT-4. Users only need to describe their commands, and then Auto-GPT will automatically generate and execute prompts step by step to achieve the expected results.

{\bfseries Code analysis with ChatGPT.} Nathalia et al.~\cite{Nathalia2023} investigated the differences between ChatGPT, novice programmers and expert programmers in solving LeetCode competition problems in terms of performance and memory efficiency. They found that ChatGPT was better than novice programmers in solving simple and medium level problems. Haque and Li~\cite{Md.2023} found that ChatGPT can identify bugs in code and provide suggestions, which not only reduces debugging time, but also makes it easier for developers to locate bugs in their programs. Tian et al.~\cite{Haoye2023} evaluated ChatGPT capabilities in accomplishing three code tasks: code generation, program repair, and code summarization. The results showed that ChatGPT performed well in these tasks, proving its potential as an assistant for programmers.

{\bfseries Worries about ChatGPT.} Even though ChatGPT has excellent performance in natural language processing and code analysis, its use raises various concerns. Lo~\cite{Lo2023} pointed out that since the training dataset of ChatGPT only goes up to 2021, it may yield biased or inaccurate knowledge when used in the education field. Moreover, the content generated by ChatGPT can bypass plagiarism detection programs, promoting a culture of student plagiarism. Chatterjee and Dethlefs~\cite{Joyjit2023} found that ChatGPT can provide answers to questions that violate legal and ethical standards by altering the way of expression. Simultaneously, they provided an example to show how ChatGPT inadvertently introduces gender and racial discrimination in the coding process. Al-Hawawreh et al.~\cite{Al-Hawawreh2023} explored how attackers can use ChatGPT to write malicious code snippets, circumvent system security measures and conduct network attacks.

\subsection{GitHub Issue Analysis}

The issues of GitHub projects have been extensively used as datasets in various research studies. Yi et al.~\cite{Yi2022} identified vulnerabilities in blockchain systems from GitHub issues and conducted a unique analysis on these vulnerabilities, namely summarizing vulnerable modules, types, and patterns. Win et al.~\cite{Win2023} identified 15 unethical behaviors based on GitHub issues and designed a tool called Etor to automatically detect these behaviors. Kallis et al.~\cite{Kallis2019} developed a tool to tag open issues in GitHub projects by analyzing their titles and text descriptions. Izadi et al.~\cite{Izadi2022} proposed a method to classify issues into bug reports, feature requests, or product support based on their text information and analyzed the priority of these issues. Huang et al.~\cite{Huang2021} collected 79 features of issues from three dimensions: clarity of issue description, complexity of changes, and skills required to address issues. After analyzing the correlation between these features and Good First Issues (GFIs), they developed a model to automatically predict GFIs for projects.

\subsection{Topic Analysis Using LDA}

LDA is a commonly used topic model in the field of software engineering. Bagherzadeh and Khatchadourian~\cite{Bagherzadeh2019} crawled posts related to big data from Stack Overflow and used LDA to identify and analyze topics related to big data, including their popularity and relevance. Haque et al.~\cite{Haque2020} created a LDA topic model on Docker related posts from the Stack Overflow community and identified the main challenges faced by developers during Docker development. Li et al.~\cite{Li2018} extracted topics from code snippets using LDA and explored the relationship between these topics and the likelihood of statements appearing in log records. Han et al.~\cite{Han2020} used LDA topic model to mine discussion topics related to three deep learning frameworks (Tensorflow, PyTorch and Theano) on Stack Overflow and GitHub platforms. Zhang et al.~\cite{Zhang2019} extracted service goals from textual descriptions of RESTful services and clustered them based on LDA results, thereby improving the accuracy of service discovery.


\section{Conclusion and Future Work} \label{sec:conclusion}


With the rapid development and widespread dissemination of ChatGPT, developers have hosted a large number of ChatGPT-related projects on GitHub. These projects have sparked extensive discussions. In this study, we investigated the categories of ChatGPT-related projects and further analyzed the popularity, difficulty, and evolution of ten issue topics discovered using the LDA topic model within three primary project categories. Based on our findings, we provide useful suggestions for project developers and hosting platforms (such as GitHub and Gitee) to better develop and manage ChatGPT-related projects. For example, the platforms can use our categories as fine-grained tags to label their projects, thus promoting the exploration experience for users. According to the popularity and difficulty of the issue topics, developers can devote more efforts and resources to solve popular and challenging issues raised for their projects. Developers can also adjust the development focus of their projects during the progress in a timely manner based on the evolution trends of issue topics. 
In future work, we plan to generate summaries of issues of a specific topic to help developers quickly understand the issues encountered by users and also recommend candidate solutions to issues by synthesizing existing solutions to similar issues, thereby accelerating project development.

\begin{acks}
This work is supported by the National Natural Science Foundation of China (62302536) and the Guangdong Basic and Applied Basic Research Foundation (2023A1515012292).
\end{acks}

\bibliographystyle{ACM-Reference-Format}
\bibliography{Refs}


\end{document}